\documentclass[aps,prl,%
 reprint,
 amsmath,amssymb,
 aps,
]{revtex4-2}
\usepackage{xcolor}
\usepackage{graphicx}
\usepackage{dcolumn}
\usepackage{bm}

\begin{document}

\preprint{APS/123-QED}

\title{Controlling isomer population using a dual-oscillator infrared free-electron laser } 

\author{Am\'erica Y. Torres-Boy*}
\affiliation{Fritz-Haber-Institut der Max-Planck-Gesellschaft, Faradayweg 4-6, 14195 Berlin, Germany}
\author{Anoushka Ghosh}
\affiliation{Fritz-Haber-Institut der Max-Planck-Gesellschaft, Faradayweg 4-6, 14195 Berlin, Germany}
\author{Myles B. T.  Osenton}
\affiliation{Fritz-Haber-Institut der Max-Planck-Gesellschaft, Faradayweg 4-6, 14195 Berlin, Germany}
\author{Akash C. Behera}
\affiliation{Fritz-Haber-Institut der Max-Planck-Gesellschaft, Faradayweg 4-6, 14195 Berlin, Germany}
\author{Sandy Gewinner}
\affiliation{Fritz-Haber-Institut der Max-Planck-Gesellschaft, Faradayweg 4-6, 14195 Berlin, Germany}
\author{Marco De Pas}
\affiliation{Fritz-Haber-Institut der Max-Planck-Gesellschaft, Faradayweg 4-6, 14195 Berlin, Germany}
\author{Heinz Junkes}
\affiliation{Fritz-Haber-Institut der Max-Planck-Gesellschaft, Faradayweg 4-6, 14195 Berlin, Germany}
\author{Wieland Schöllkopf}
\affiliation{Fritz-Haber-Institut der Max-Planck-Gesellschaft, Faradayweg 4-6, 14195 Berlin, Germany}
\author{Alexander Paarmann}
\affiliation{Fritz-Haber-Institut der Max-Planck-Gesellschaft, Faradayweg 4-6, 14195 Berlin, Germany}
\author{Gert von Helden}
\affiliation{Fritz-Haber-Institut der Max-Planck-Gesellschaft, Faradayweg 4-6, 14195 Berlin, Germany}
\author{Gerard Meijer*}
\affiliation{Fritz-Haber-Institut der Max-Planck-Gesellschaft, Faradayweg 4-6, 14195 Berlin, Germany}

\date{\today}

\begin{abstract}
We report on the control and characterization of the isomer population of ions inside superfluid helium nanodroplets, using two-color operation of a dual-oscillator infrared free-electron laser. The timing of both lasers is highly synchronized and their frequencies (or "colors") can be tuned independently over a wide range. Interaction of the singly deuterated proton-bound dimer of dihydrogen phosphate and formate inside helium nanodroplets with both colors enables the control over its isomer population and the recording of - one-color hidden - infrared spectra of individual isomers.
\end{abstract}

\maketitle



Chemical transformations can be described by the dynamics of molecular systems on their potential energy surfaces (PESs), determining the path of a reaction. Its rate is determined by the shape of the PES and possible barriers separating reactants and products. Empirically and in a simple form, this was captured already at the end of the 19th century in the famous Arrhenius equation by introducing an activation barrier for a reaction that needs to be overcome.\cite{arrhenius1889reaktionsgeschwindigkeit} These days, sophisticated theoretical frameworks exist to predict reaction dynamics and rate constants. For their application, however, approximations need to be made for all but the very simplest systems.  Gas-phase studies are relevant in this context as these can provide molecular-level insight, unperturbed by the environment.  A well-defined amount of energy can be deposited into molecular systems via photons, enabling controlled studies of intramolecular energy flow and reactivity in isolated systems.\cite{levine2005molecular} 

Starting from cold molecular systems, absorption of infrared (IR) photons effectively heats the molecule, and the subsequent processes can be probed under microcanonical conditions.
IR photons selectively excite vibrational modes, thereby initiating thermally driven processes such as conformational changes\cite{dian2002peptidedynamics, maccoas2004IRinduc, douberly2007IRdoubleresIsom, dian2008MWmesIRiso} and proton transfer,\cite{clarkson2005laserinitshuttl,rana2025microcanonical} processes that are governed by non-covalent interactions. During the past decades, these processes have been extensively investigated using IR or combined IR–UV excitation schemes; structural changes are initiated by one laser pulse, after which the resulting products are probed. Experiments have been conducted on neutral systems in a variety of environments, including molecular beams\cite{dian2002peptidedynamics, gerhards2018isomerspecIRUV,moon2025IRUV} and helium nanodroplets,\cite{douberly2005IRIRphotoisomerization, douberly2007IRdoubleresIsom} as well as on ionic molecular systems confined in ion traps studying, for instance, biomolecules\cite{leavitt2012isomerspecificIRIR, seaiby2016IRconfIontrap} and water clusters.\cite{khuu2023EigenZundelisovib, rana2025onset}
 Using IR-induced population-transfer spectroscopies, structural rearrangements in a dipeptide\cite{dian2002peptidedynamics} and the kinetics of proton transfer in a microhydrated organic compound have been directly monitored.\cite{rana2025microcanonical}

Infrared free-electron lasers (IR-FELs) have proven to be ideal tools for different variants of background-free ("action") spectroscopy on neutral and charged species.\cite{oomens2006IRMPDreview, polfer2009ReviewIonsbio,FlorezHedrop15, javsikova2018IRMPDwithFELs} Although a variety of tunable \textcolor{black}{table-top} IR sources are commercially available, none of these have the wide tuning range that IR-FELs offer. In particular, the access at long wavelengths ($\lambda \geq$ 15 $\mu$m) remains limited with commercial sources. Moreover, IR-FELs with a high micro-pulse repetition rate ({\it vide~infra}), like FELIX in The Netherlands, CLIO in France, and the IR-FEL at the Fritz Haber Institute (FHI-FEL), provide a very large number of photons over an extended period of time. This can be used to advantage to deposit large amounts of energy into (gas phase) samples via multiple excitation/relaxation cycles.

Here, we employ the advantages of ion-based IR spectroscopy with the well-defined conditions of superfluid helium nanodroplets. The helium nanodroplets provide a dissipative and isothermal environment at 0.37~K, leading to rapid cooling of the embedded ions after IR excitation.  Together with the long interaction time (up to 10~$\mu$s) of the doped helium droplets with the IR-FEL, complete population transfer between two distinct isomers can be induced via multiple excitation/relaxation cycles.  Using a novel two-color dual-oscillator IR-FEL, the isomer population can be fully controlled, and clean spectra of the individual isomers can be recorded.

The temporal structure of the light of an FEL is determined by the temporal structure of the electron beam from which the light is generated. The generation of two IR-frequencies from a single electron beam in an optical cavity was first demonstrated at CLIO in 1994. In that setup, a step-tapered undulator was used,\cite{jaroszynski1994PRLtwocolor} but this only provides a very limited tunability.\cite{prazeres1998twocolourFELCLIO,ortega2003twoIRexp}
An approach to achieve true two-color IR-FEL operation is to form two beams of electrons from one and to couple these into two separate optical cavities with independently tunable undulators. By spatially separating alternate electron bunches from an RF-driven linear accelerator (linac), perfect synchronization of the two electron beams is guaranteed by the highly coherent RF fields of the linac.\cite{schwettman1989two}

The electron gun and RF-driven linac of the FHI-FEL generates 10~$\mu$s-long electron bunch trains (macro-bunches) at a 10~Hz repetition rate. Each macro-bunch is composed of ps-duration bunches of electrons, spaced by 1~ns, that is, the electron bunches come at a 1~GHz repetition rate. A 500~MHz "kicker" cavity installed downstream of the accelerator can deflect the electron bunches alternately $\pm2^\circ$ to the left or to the right, thus creating two spatially separated, synchronized electron bunch trains, each with the bunches separated by 2~ns (500 MHz bunch repetition rate). These macro-bunches are coupled into two oscillator FELs, referred to as the mid-IR and far-IR FEL, each one including a variable-gap undulator, located  within 5.4-meter-long optical cavities. The output of the mid-IR FEL consists of 10~$\mu$s duration macro-pulses of light that can be tuned from 2.9 - 50~$\mu$m, with up to 100~mJ of pulse energy; the far-IR FEL can be tuned from 4.5 - 160~$\mu$m, and delivers up to 150~mJ. A detailed technical description of the dual-oscillator setup with a full characterization of both IR-FELs will be given elsewhere.\cite{wieland2026} \\

 In our experimental setup, the nanodroplet beam passes through an ion trap, capturing mass-selected positive or negative ions that are stored in the trap, as described elsewhere.\cite{FlorezHedrop15} The speed of the nanodroplets in the beam is typically 500~m/s, and when the nanodroplets are heavy enough, their kinetic energy is sufficient that the embedded ions - and only those - can overcome the trapping potential. This yields a directed beam of mass-selected, cryogenically cooled ions inside helium nanodroplets, that is, a well-defined sample that can subsequently be spectroscopically interrogated.\cite{zhang2012IRionshedrop, raston2013formaldehydevibrex, FlorezHedrop15,verma2019IRHeD, singh2025IRNH3}
When the ion inside the helium nanodroplet absorbs an IR photon, it is transiently heated and redistributes its internal energy, which is efficiently transferred to the surrounding helium droplet. The droplet dissipates this energy through evaporative cooling and concomitant shrinking, with approximately one helium atom evaporated for every 5~cm$^{-1}$ of energy deposited.\cite{lehmann1998Hespecmatrix, toennies2004review}  This allows the next IR photon absorption to  take place again from the ground state of the ion, now embedded in a somewhat smaller droplet. This excitation/relaxation process can be repeated many times during the interaction with the IR radiation. By measuring the amount of signal on the mass-to-charge channel of the bare ion as a function of the frequency of the IR-FEL, vibrational spectra displaying narrow bands, have thus been recorded for a wide variety of systems.\cite{FlorezHedrop15,Mucha20017-Glycan,Thomas2019-CAFluo, FP2024}\\

When isomerization occurs in ionic complexes embedded in helium nanodroplets, the excitation/relaxation cycle described above can be interrupted, leading to the absence of spectral lines. When transitions of the different isomers coincidentally overlap, it can result in the spurious appearance of ion signal. It has been speculated that such IR-induced isomerization is the reason for the unexpected spectral structure observed in the study on the proton-bound dimer of dihydrogen phosphate and formate.\cite{FP2024} In those experiments, the observed spectra of the fully hydrogenated and of the fully deuterated complexes agree well with the calculated ones, whereas in the spectra of the partially deuterated species, many lines appear to be absent.  \\

\begin{figure}[t]
  \includegraphics[width=87mm]{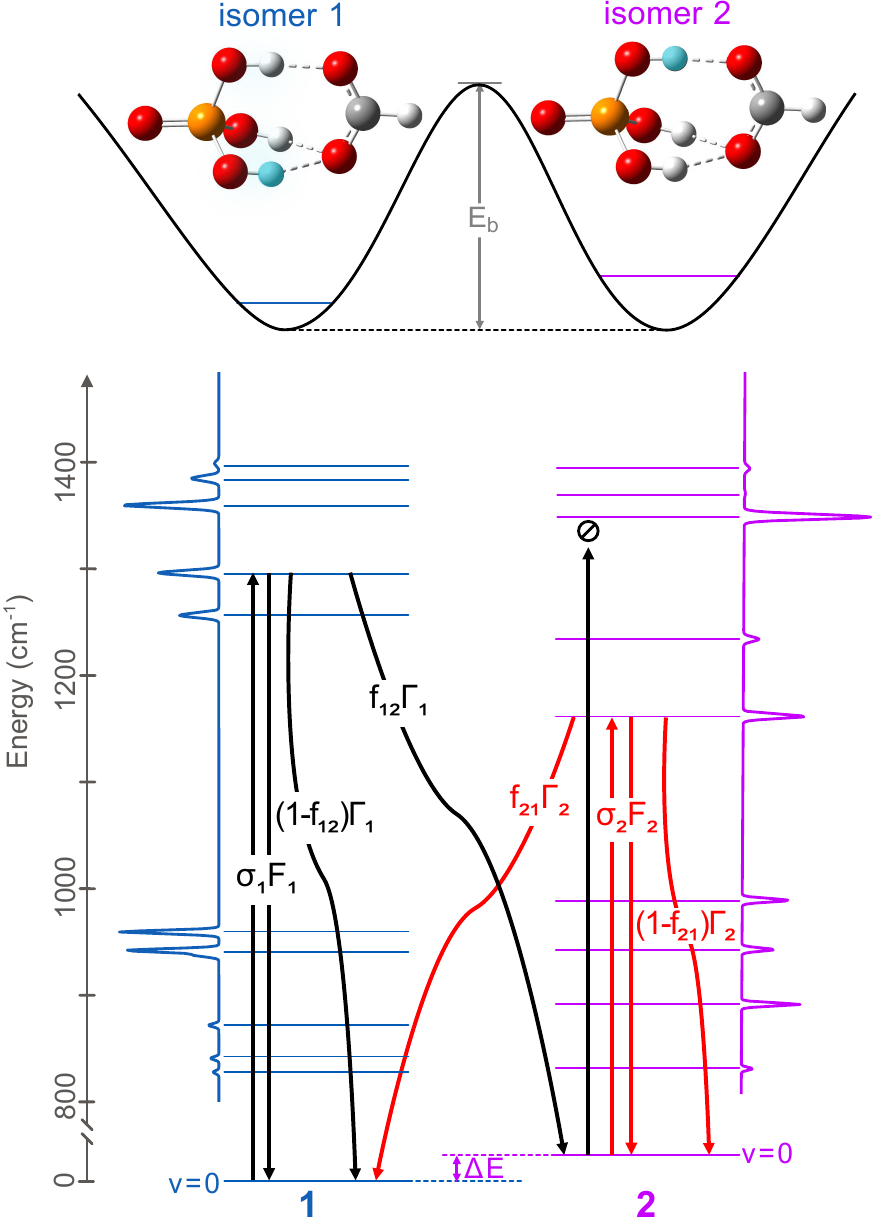}
  \caption{Scheme of the vibrational energy levels for the two possible singly deuterated isomers of the proton-bound dimer of dihydrogen phosphate and formate that can be reached via IR allowed transitions from the vibrational ground state. The calculated, distinct IR spectra for both isomers are shown along the vertical energy axis, pointing to the left for isomer~1 and to the right for isomer~2. The geometric structures of both isomers are shown on top. The excitation, stimulated emission, and relaxation rates that are relevant when one IR-FEL (processes indicated in black) and both IR-FELs (adding the processes indicated in red) are used are indicated.}
  \label{fig:levelscheme}
\end{figure}

In the phosphoric acid–formate complex,  a singly negatively charged ion, the two moieties are connected by three hydrogen bonds. The upper part of Fig.~\ref{fig:levelscheme} shows the structures of the singly deuterated species, in which two H atoms (white) and one D atom (blue) participate in the hydrogen-bond network. Two distinct isomers exist: one in which the D atom occupies a position that is symmetry-equivalent to that of an H atom (isomer~1, C$_{1}$ symmetry), and one in which the D atom occupies a unique position (isomer~2, C$_{s}$ symmetry). The energy difference $\Delta E$ between these isomeric structures is solely determined by the zero-point vibrational energy difference. Isomer 1 is calculated to be (only) about $\Delta E$ = 17~cm$^{-1}$ lower in energy than isomer~2.\cite{SM} Both structures are therefore expected to be present in the liquid nitrogen-cooled ion trap that the helium nanodroplets pass through. Given that isomer~1 comes in two symmetry-equivalent versions, its abundance in the ion trap will be about a factor of two higher than that of isomer~2. Shock-freezing into helium nanodroplets can preserve this ratio, and both isomers are expected to contribute to the vibrational spectrum.\\

The vibrational modes of both isomers in the 700~-~1300~cm$^{-1}$ region involve extensive motion of the hydrogen-bonded H and D atoms, leading to distinct mid-IR spectra. This is pictorially shown in Fig.~\ref{fig:levelscheme} where the calculated mid-IR spectra of isomer~1 (left) and isomer~2 (right) are shown along a vertical energy axis. The energy levels that can be reached in our experiments via IR-allowed transitions from the ground state are drawn in as solid horizontal lines. In this figure, the various excitation and decay processes that are relevant when an IR-FEL is resonant with a transition in isomer~1 are indicated in black. The resonant excitation and stimulated emission rate $\sigma_1 F_1$, with $\sigma_1$ the absorption cross-section and $F_1$ the laser fluence (in photons/(cm$^2$s)), as well as the overall decay rate of the excited level $\Gamma_1$ (s$^{-1}$), are shown. As the barrier between the isomers is calculated to be approximately 357~cm$^{-1}$,\cite{SM} which is lower in energy than the levels shown in Fig.~1, a fraction of the decay rate, given by $0 \leq f_{12} \leq 1$, can lead to interconversion, thereby increasing the population of isomer~2. When the IR-FEL is not resonant with a transition in isomer~2, as schematically depicted in black in Fig.~\ref{fig:levelscheme}, all population will ultimately accumulate in isomer~2, and no further absorption of photons will occur. When a second IR-FEL that is resonant with a transition in isomer~2 is used, the various processes indicated in red in Fig.~\ref{fig:levelscheme} take place as well. The fractional decay rate $f_{21} \Gamma_2$ from the excited level in isomer~2 back to isomer~1 closes the excitation/relaxation cycle and makes repeated absorption of photons by both IR-FELs possible. \\

For the ion detection process, the total amount of energy, E$_{tot}$, that is deposited after sequential absorption of multiple IR photons is the crucial parameter. We have observed that, with the distribution of droplet sizes that we have in the experiment, the ion signal scales in a good approximation linearly with E$_{tot}$. When using a single IR-FEL that is resonant with a transition from the ground state to a level with energy $E_i$ in isomer $i$ (for $i$=1,2), all population will ultimately be transferred to the other isomer. For an interaction time $T$ of the laser with the helium nanodroplets that is long enough, this steady state will be reached. In the following, we treat the system as interacting with a continuous light pulse of duration $T$.  When the IR-FEL is resonant with isomer~1, an energy E$_{tot}$ given by 
\begin{equation}
E_{tot} = \sum_{n=0}^{\infty} (1-f_{12})^n \cdot [E_1 - f_{12} \cdot \Delta E] \approx \frac{1}{f_{12}} \cdot E_1
\end{equation}
is deposited in those helium nanodroplets that originally contained isomer~1, i.e., in about two-thirds of the anion-doped helium nanodroplets. Similarly, when the single IR-FEL is resonant with a transition in isomer~2, a total energy of E$_{tot} \approx E_2/f_{21}$ is deposited in about one-third of the anion-doped droplets.\\

When a second IR-FEL is added that is resonant with a transition in the other isomer, the excitation/relaxation cycle is closed. The steady state solution of the rate equations yields the fractional population $n^*_1$ and $n^*_2$ in the excited vibrational level of either isomer. The total energy that is deposited in the helium nanodroplet is then given by
\begin{equation}
E_{tot} = n^*_1 \cdot (E_1 - f_{12} \cdot \Delta E) \cdot \Gamma_1 \cdot T + n^*_2 \cdot (E_2 + f_{21} \cdot \Delta E) \cdot \Gamma_ 2 \cdot T
\end{equation}
which is found as 
\begin{equation}
E_{tot} = \frac{\sigma_1 F_1 \cdot \sigma_2 F_2 \cdot (f_{21} \cdot E_1 + f_{12} \cdot E_2) \cdot T}{\sigma_1 F_1 \cdot f_{12} \cdot[1 + \frac{2 \cdot \sigma_2 F_2}{\Gamma_2}] + \sigma_2 F_2 \cdot f_{21} \cdot[1 + \frac{2 \cdot \sigma_1 F_1}{\Gamma_1}]}
\end{equation}
and this amount of energy is deposited in all anion-doped helium nanodroplets. When the relaxation is fast relative to the IR excitation, i.e., when $\Gamma_i$ $>>$ $\sigma_i F_i$ for $i$=1,2, the terms in square brackets in the denominator can be set equal to one. The optimum value for $E_{tot}$ is obtained when $f_{12} \cdot \sigma_1 F_1$ equals $f_{21} \cdot \sigma_2 F_2$, that is, when the IR-excitation induced $1 \rightarrow 2$ and $1 \leftarrow 2$ isomerization rates are balanced. In that case, the expression for $E_{tot}$ reduces to
\begin{equation}
E_{tot} \approx \frac{1}{2} \cdot [\sigma_1 F_1 \cdot T \cdot E_1 + \sigma_2 F_2 \cdot T \cdot E_2]
\end{equation} 
The product $\frac{1}{2} \cdot \sigma_i F_i \cdot T$ (for $i$=1,2) is the total number of excitation/relaxation cycles induced by the IR-FEL that is in resonance with isomer~$i$ during its $\frac{1}{2}T$ interaction time, i.e. it is the number of IR photons that this IR-FEL deposits in the helium nanodroplets.   \\

\begin{figure}[t]
  \includegraphics[width=80mm]{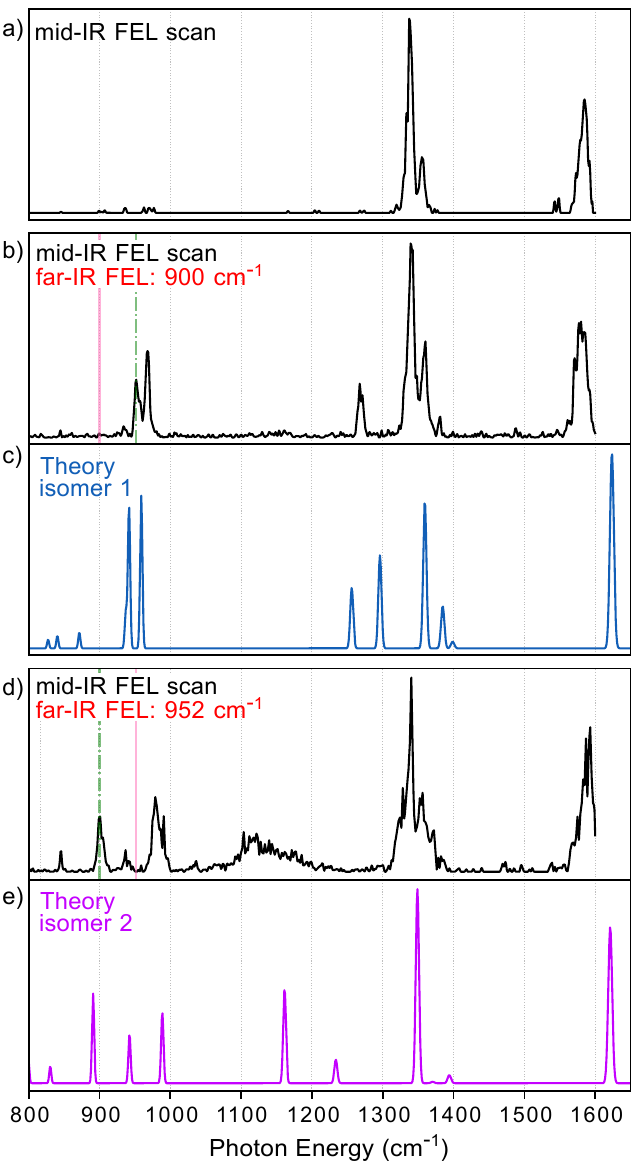}
  \caption{IR spectra of the singly deuterated proton-bound dimer of dihydrogen phosphate and formate in helium nanodroplets, obtained by recording the signal of the bare anion as a function of the frequency of the mid-IR FEL. In the upper panel, a), only the mid-IR FEL is used. In panels b) and d), the far-IR FEL is kept fixed at 900~cm$^{-1}$ and 952~cm$^{-1}$, respectively, while scanning the mid-IR FEL. The spectra calculated at the B3LYP-D3(BJ)/aug-cc-pV(T+d)Z level of theory in the harmonic approximation of isomers 1 and 2 are displayed in panels c) and d) for comparison.}
  \label{fig:spectra}
\end{figure}

In Fig.~\ref{fig:spectra}, different IR spectra of the singly deuterated proton-bound dimer of dihydrogen phosphate and formate in helium nanodroplets are shown in the 800 - 1600~cm$^{-1}$ region. The spectra are recorded by scanning the frequency of the mid-IR FEL while recording the signal intensity at the mass-to-charge ratio of the bare anion. In recording the spectrum shown in panel (a), the mid-IR FEL is the only laser used, and ion signal is only observed at wavelengths where the vibrational signatures of both isomers coincidentally overlap within the bandwidth of this laser.
The spectra shown in panels (b) and (d) are recorded by scanning the mid-IR FEL while the far-IR simultaneously interacts with the anionic complexes in the helium nanodroplets. The far-IR FEL is kept at a fixed frequency, overlapping with a vibrational band that is unique for either one of the isomers. The latter are found in an iterative process, stepping the frequency of the far-IR FEL through the  800-1000 cm$^{-1}$ spectral region.
With the far-IR FEL fixed at 900~cm$^{-1}$, a completely new spectrum is observed (panel (b)), which is in good agreement with the calculated spectrum of  isomer~1 (panel (c)). When the far-IR FEL is kept fixed on a resonance of isomer~1 at 952~cm$^{-1}$, the mid-IR scan yields yet another distinct IR spectrum (panel (d)), which resembles the calculated spectrum of isomer~2 (panel (e)). Note that the spectrum  of isomer~2 indeed has a unique resonance at 900~cm$^{-1}$. These measurements demonstrate that the far-IR FEL, when in resonance with only one isomer, efficiently depletes the population of that isomer, enabling the recording of the spectrum of the other isomer with the spatially and temporally overlapping mid-IR FEL. For the assignment of the various modes that are present in the calculated spectra of both isomers, we refer to our earlier study.\cite{FP2024} We emphasize that the corresponding experimental spectra could not be obtained then, i.e., when using a single IR-FEL, also not with an increased laser bandwidth and at the highest laser fluence used. \\

 The microcanonical temperature that the complex reaches after the absorption of a 1000~cm$^{-1}$ photon is estimated to be about 300~K.\cite{SM} Therefore, our experiments actually probe isomerization in the thermal regime.   Following IR excitation, intramolecular vibrational redistribution (IVR) will take place. In a study on formaldehyde in helium nanodroplets IVR is proposed to occur faster than 20~ps.\cite{raston2013formaldehydevibrex} Following IVR, two competing processes can take place: isomerization of the excited species and vibrational relaxation, i.e. cooling down to the ground state.  The latter is expected to occur at comparable timescales for both isomers, faster than nanoseconds.\cite{zhang2012IRionshedrop} Our experiments clearly show that isomerization does occur; therefore, isomerization is significantly faster than vibrational relaxation. But, even in this case, the back isomerization would be fast as well. Consequently, $f_{12}$ cannot be equal to one, but is limited by statistics. 
A microcanonical RRKM rate analysis for the isomerization process is reported in the supplementary material.\cite{SM}  The reaction rate constant of such a process is calculated to be in the range of 10$^{11}$-10$^{12}$~s$^{-1}$, corresponding to isomerization on the timescale of a few picoseconds.\\

 It is seen from the spectrum shown in panel (b) of Fig.~\ref{fig:spectra} that isomer~1 has a relatively strong and spectrally isolated mode at 1270~cm$^{-1}$. Excitation with one IR FEL on this mode while scanning the other enables the full recovery of the spectral structure of isomer~2.\cite{SM} In hindsight, this mode at 1270~cm$^{-1}$ can also be recognized just above the noise level in panel (a) of Fig.~\ref{fig:spectra},  almost a factor of a hundred weaker than the main peak in that spectrum. Given that two-thirds of the anion-doped nanodroplets can contribute to the signal at 1270~cm$^{-1}$, we conclude from Eq. (1) and (4) that the factor $f_{12}^{-1}$ is at least a factor 50 smaller than the number of IR photons that are absorbed when the excitation/relaxation cycle is closed. As we estimate that we typically need to absorb several hundred IR photons to record a good signal, it follows that $f_{12}$ for this particular mode might be close to the statistical limit of 1/3; the statistical limit of $f_{21}$ is twice as large.\\

In conclusion, two-color operation of a dual-oscillator IR-FEL is demonstrated to enable control over the isomer population of ionic complexes in helium nanodroplets. The large amount of energy in the many microseconds duration of the macro-pulse of the IR-FEL, combined with the fast cooling rates provided by the helium environment, allows for the complete conversion of one isomer to the other. This allows the isomer-selective recording of spectra with a dual-oscillator IR-FEL, spectra that are unable to be obtained with a single IR excitation source. The IR-induced isomerization will depend on the nature and energy of the excited modes, in particular for modes that are relatively close to the isomerization barrier, allowing for detailed studies of the dynamics. For the two-color IR-FEL experiments reported here, temporal overlap of the macro-pulses from each of the FELs has been sufficient, and the precise timing between the micro-pulses from both IR-FELs has been uncritical. Synchronization of the micro-pulses on the ps-level is inherent in the present design, however, and offers great prospects for future studies of dynamical processes.\cite{wieland2026} \\

\section*{Acknowledgments}
\begin{acknowledgments}
A.Y.T.B. acknowledges support from the IMPRS for Elementary Processes in Physical Chemistry. A.G. acknowledges the DAAD-RISE funding. A.C.B. acknowledges funding from the Max Planck-Radboud University Center for Infrared Free-Electron Laser Spectroscopy.
\end{acknowledgments}

\section*{Data Availability Statement}

The data that support the findings of this study are available
from the corresponding authors upon reasonable request.

\bibliography{endnote}

\end{document}